\documentclass[review]{elsarticle}
\usepackage{float}
\usepackage{geometry}
\usepackage{epstopdf}
\usepackage{url}
\usepackage{hyperref}
\usepackage{color}
\usepackage{amsmath}

\journal{COMMUN NONLINEAR SCI}

\begin{document}

\begin{frontmatter}

\title{Comparative analysis of time irreversibility and amplitude irreversibility based on joint permutation}

\author[mainaddress,adaddress]{Wenpo Yao\corref{mycorrespondingauthor}}
\cortext[mycorrespondingauthor]{Corresponding author}
\ead{yaowp@njupt.edu.cn}
\author[firstadress]{Wenli Yao}
\author[Tirdadress]{Rongshuang Xu}
\author[mainaddress]{Jun Wang\corref{mycorrespondingauthor}}
\ead{wangj@njupt.edu.cn}

\address[mainaddress]{School of Geographic and Biologic Information, Smart Health Big Data Analysis and Location Services Engineering Lab of Jiangsu Province, Nanjing University of Posts and Telecommunications, Nanjing 210023, China}
\address[adaddress]{Key Laboratory of Computational Neuroscience and Brain-Inspired Intelligence (Fudan University), Ministry of Education}
\address[firstadress]{Department of Hydraulic Engineering, School of Civil Engineering, Tsinghua University, Beijing 100084, China}
\address[Tirdadress]{School of Applied Meteorology, Nanjing University of Information Science \& Technology, Nanjing 210044, China}

\begin{abstract}
Although time irreversibility (TIR) and amplitude irreversibility (AIR) are relevant concepts for nonequilibrium analysis, their association has received little attention. This paper conducts a systematic comparative analysis of the relationship between TIR and AIR based on statistical descriptions and numerical simulations. To simplify the quantification of TIR and AIR, the amplitude permutation and global information of the associated vector are combined to produce a joint probability estimation. Chaotic logistic, Henon, and Lorenz series are generated to evaluate TIR and AIR according to surrogate theory, and the distributions of joint permutations for these model series are measured to compare the degree of TIR and AIR. The joint permutation TIR and AIR are then used to investigate nonequilibrium pathological features in epileptic electroencephalography data. Test results suggest that epileptic brain electrical activities, particular those during the onset of seizure, manifest higher nonequilibrium characteristics. According to the statistical definitions and targeted pairs of joint permutations in the chaotic model data, TIR and AIR are fundamentally different nonequilibrium descriptors from time- and amplitude-reversibility, respectively, and thus require different forms of numerical analysis. At the same time, TIR and AIR both provide measures for fluctuation theorems associated with nonequilibrium processes, and have similar probabilistic differences in the pairs of joint permutations and consistent outcomes when used to analyze both the model series and real-world signals. Overall, comparative analysis of TIR and AIR contributes to our understanding of nonequilibrium features and broadens the scope of quantitative nonequilibrium measures. Additionally, the construction of joint permutations contributes to the development of symbolic time series analysis.
\end{abstract}


\end{frontmatter}


\section{Introduction}
Time reversibility describes the property whereby a process is invariant under reverse time scale, whereas amplitude reversibility describes the invariant probabilistic properties of a process with respect to amplitude
reversal~\cite{Yao2020CNS}. The irreversibility metrics, namely time irreversibility (TIR) and amplitude irreversibility (AIR), are fundamental properties of nonequilibrium systems~\cite{Weiss1975,Ropke2013,Costa2005,Fang2019}, and serve as reliable 
parameters for the determination of nonlinearity, a necessary condition for chaotic behavior. If the fluctuations in real-world systems are not symmetric, the related time series is intuitively irreversible. The importance of reversibility has long been illustrated by the standard modeling paradigm in complex dynamical systems analysis, such as physiological conditions, financial 
phenomena, and meteorological activities~\cite{Yao2020CNS,Costa2005,Fang2019,Lynn2021,Wan2018,Martin2018}.

Statistically, TIR and AIR can be quantified by calculating the joint probabilistic differences between the original and reversed series or between the symmetric vectors of the series~\cite{Yao2020CNS,Wan2018,Martin2018,Yao2020ND,Martin2019,Zanin2021C}. 
However, both of these approaches are nontrivial. In practice, if the data can be represented by known parametric distributions (e.g.,~Gaussian), model-based estimators could be employed as functions of the parameters to measure the data probability distributions; otherwise, model-free probability estimators (e.g.,~the kernel and nearest-neighbor estimators~\cite{Rojo2018,Marina2007,Schreiber2000T}) are more appropriate, particularly for complex systems with unknown 
characteristics~\cite{Xiong2017,Hlavac2007}. These traditional estimators generally reconstruct time series in which the symmetric or corresponding vectors might not match. Because of the limitations of these probability estimations, TIR and AIR are generally quantified by simplifying the time series of interest, i.e.,~coarse-grained approaches. For instance, 
Costa~\cite{Costa2005} calculated the difference between the average energy of activation and relaxation, and then simplified the metric using up--down probabilistic difference~\cite{Costa2008}, as employed in the Guzik~\cite{Guzik2006}, Porta~\cite{Porta2008}, and Cammarota indexes~\cite{Cammarota2007}. Symbolic time series analysis~\cite{Daw2003,Meyer2021,Raul2021} has gained popularity because of its computational effectiveness, robustness to noise, and simplified statistical analysis. Visibility irreversibility~\cite{Lacasa2008,Luque2009} can be detected by measuring the distinguishability between the in--out degree distributions~\cite{Lacasa2012,Lacasa2015,Flanagan2016,Donges2013}. Among these 
simplified approaches, those based on permutations are particularly popular. Ordinal patterns do not impose any model assumptions and inherit the structural dynamics of time series~\cite{Bandt2002,Bandt2020,Zanin2021P}, and their construction is highly relevant to the estimation of joint probabilities~\cite{Martin2018,Yao2020ND,Martin2019,Zanin2021C}. Note that the 
asymmetric indexes proposed by Costa, Guzik, and Porta~\cite{Costa2008,Guzik2006,Porta2008} are a special case of permutation 
irreversibility considering only neighboring double values. However, several issues might affect the permutation TIR or AIR. Equal values change the structure and probability distributions of ordinal patterns, potentially leading to significantly different or even contradictory results~\cite{Yao2019E,Bian2012,Zunino2017,Yao2020APL}. Equal-values vectors 
might have a self-symmetric structure describing some specific physical implication, i.e.,~time-reversibility. The existence of forbidden permutations means that division-based parameters (e.g.,~the Kullback--Leibler distance) should not be directly applied in quantitative TIR and AIR~\cite{Yao2020ND,Yao2019Sym}. Moreover, the two basic ordinal patterns, i.e.,~ original and amplitude permutations~\cite{Yao2022PLA}, are different. Original permutation consists of the indexes of reorganized values in the 
original vector, and is not a direct reflection of the amplitude structure of the vectors. This has no effect if permutations are used as labels of different vectors, such as for permutation entropy, synchronization, or network analysis~\cite{Bandt2002,Bandt2020,Zanin2021P,Zou2019,Pessa2020}; however, if the original permutation is used as a temporal replacement for the vector in quantifying TIR, it might yield misleading results or even conceptual errors~\cite{Yao2020CNS,Yao2019Sym}. These coarse-grained approaches simplify the estimation of joint probability distributions and enable the quantification of TIR and AIR, although in practical applications, the corresponding relationship between the simplified forms and vectors requires particular attention.

Through comparative analysis of the vector structure and its permutation, the present authors have previously corrected the conceptual error underlying the original symmetric permutations used to quantify TIR~\cite{Yao2020ND,Yao2019Sym,Yao2022PLA}, and derived the concept of amplitude-reversibility based on the symmetric permutations~\cite{Yao2020CNS}. The nonequilibrium descriptors provided by TIR and the novel AIR are linked---they are different theoretical concepts, but experimental analysis has shown that they produce consistent findings~\cite{Yao2020CNS,Yao2020ND,Yao2019Sym}. However, their relationship and its implications have not been studied. In particular, the following issues remain unresolved: (1) TIR and AIR are fundamentally different concepts and target different aspects of nonequilibrium systems, among which AIR is derived from the conceptual mistakes of quantitative TIR. Therefore, what can we learn about the quantification of nonequilibrium processes from their conceptual and numerical differences? (2) TIR and AIR exhibit a high level of consistency in terms of their theoretical definitions and numerical signal analysis results. What are the reasons for these similarities and do they contribute to the analysis of nonequilibrium systems?  (3) Permutation analysis plays an important role in quantifying TIR and AIR~\cite{Yao2020CNS,Martin2018,Yao2020ND,Martin2019,Zanin2021C}. Original and amplitude permutations both convey basic vector structures~\cite{Yao2022PLA}, but could they be further improved for TIR and AIR analysis? Overall, the associations among TIR and AIR, and their quantification based on ordinal patterns and physical implications, require further study.

This paper explores the relationships between TIR and AIR based on joint ordinal patterns. The concepts of TIR and AIR are first introduced from a statistical perspective. Further, to simplify the probability estimation for quantifying TIR and AIR, the two basic ordinal patterns are distinguished, and the joint permutation is constructed as an alternative to the raw vector. In joint permutation TIR and AIR, the influences of equal values and forbidden permutations are elucidated. Model series and their surrogate data are generated for numerical analysis, allowing us to test whether TIR and AIR hold. At the same time, the paired joint permutations and their probability distributions are clarified for the comparison of TIR and AIR. Joint permutation TIR and AIR are then applied to brain electrical signals obtained from epileptic patients, and the nonequilibrium features of these real-world physiological signals are detected. Finally, several issues related to the permutation TIR and AIR and their applications to epileptic brain signals are discussed.

\section{Methodology}
TIR is an important characteristic of nonequilibrium systems, and AIR is a corresponding parameter from the perspective of amplitude reversal. The quantification of TIR and AIR requires the calculation of joint probabilistic differences, which is not trivial. Most quantitative measures are based on coarse-grained approaches, such as the up--down differences~\cite{Costa2008,Guzik2006,Porta2008,Cammarota2007} or symbolic motifs~\cite{Daw2003,Raul2021}, among which ordinal pattern is particularly popular~\cite{Bandt2002,Bandt2020,Zanin2021P}.

\subsection{Statistical definitions of TIR and AIR}
A process is defined as time-reversible if it is invariant under the reversal of time scale~\cite{Weiss1975,Yao2020ND,Kelly1979} and as amplitude-reversible if it is invariant with respect to amplitude reversal~\cite{Yao2020CNS}. Let us first introduce their statistical definitions.

Statistically, if a process and its time-reversal form have the same joint probability distributions, it is time-reversible; otherwise, it is time-irreversible. Weiss~\cite{Weiss1975} defined a stationary process $X(t)$ as time-reversible if $\{X(t_{1}),X(t_{2}), \ldots,X(t_{m})\}$ and $\{X(-t_{1}),X(-t_{2}), \ldots,X(-t_{m})\}$ have the same joint probability distributions for all $t_{1},t_{2}, \ldots,t_{m}$ and $m$; otherwise, $X(t)$ is time-irreversible. Alternatively, if the process is time-reversible, it is also temporally symmetric and so the associated multidimensional vectors and their symmetric forms have the same joint probabilities. According to Kelly~\cite{Kelly1979}, if $X(t)$ is time-reversible, $\{X(t_{1}),X(t_{2}), \ldots,X(t_{m})\}$ and $\{X(-t_{1}+n),X(-t_{2}+n), \ldots,X(-t_{m}+n)\}$ have the same joint probability distributions for every $t_{1},t_{2}, \ldots,t_{m}$ and $n$. Under this definition, letting $n=t_{1}+t_{m}$, time-reversibility implies that the symmetric vectors $\{X(t_{1}),X(t_{2}), \ldots,X(t_{m})\}$ and $\{X(t_{m}), \ldots,X(t_{2}),X(t_{1})\}$ have the same joint probability  distributions. Therefore, TIR is also defined as temporal asymmetry~\cite{Costa2005,Costa2008,Guzik2006,Porta2008,Diks1995}.

Weiss~\cite{Weiss1975} imposes the condition of stationarity in definition, and provides examples of stationary processes that are time-irreversible, such as shot noise processes. Kelly~\cite{Kelly1979} proves that time-reversibility implies that a reversible process is stationary. The two definitions suggest that time-reversible processes are a subset of stationary ones, i.e.,~time-reversibility implies stationarity, but the reverse is not true~\cite{Yao2020ND}. However, Ramsey and Rothman~\cite{Ramsey1995} demonstrate that it is possible to generate time-reversible processes that are nonstationary, suggesting that time-reversibility and stationarity are separate concepts and neither implies the other.

Corresponding to the time reversibility, amplitude reversibility describes a process that has same properties as its amplitude-reversal form. From a statistical perspective, amplitude reversibility is defined as follows: for a zero-mean process $X(t)=X(t)-\mu$, if $\{X(t_{1}),X(t_{2}), \ldots,X(t_{m})\}$ and its amplitude reverse $\{-X(t_{1}),-X(t_{2}), \ldots,-X(t_{m})\}$ have same joint probability distributions for every $t_{1},t_{2}, \ldots,t_{m}$ and $m$, $X(t)$ is amplitude reversible; otherwise, $X(t)$ is amplitude irreversible~\cite{Yao2020CNS}.

\begin{figure}[htb]
	\centering
	\includegraphics[width=8.3cm,height=6cm]{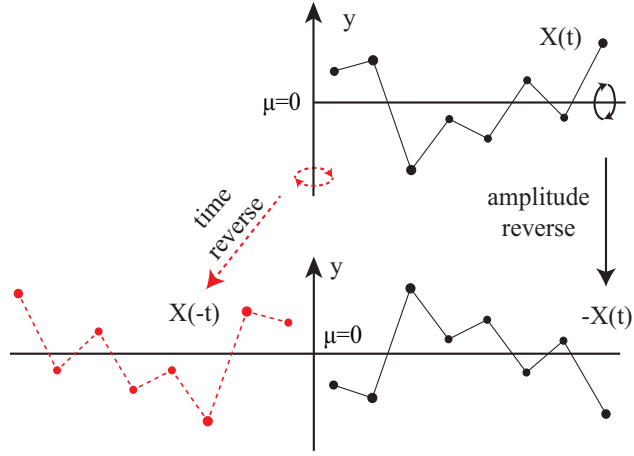}
	\caption{An original series and its time-reversible and amplitude-reversible sequences. The series $X(t)$ is a zero-mean process $\mu$=0. Red dashed time-reverse $X(-t)$ of $X(t)$ is on the left side of $y$-axis; the amplitude-reverse $-X(t)$ is on the right side of $y$-axis.}
	\label{fig1}
\end{figure}

According to the conceptual definitions, TIR involves calculating the joint probabilistic differences between $X(t)$ and its time-reversal $X(-t)$, whereas AIR measures those of $X(t)$ and its amplitude-reversal $-X(t)$.  Fig.~\ref{fig1} illustrates the original $X(t)$ and its time-reversal $X(-t)$ and amplitude-reversal $-X(t)$.

The statistical definitions and Fig.~\ref{fig1} show that TIR and AIR are fundamentally different concepts and measure different characteristics of nonequilibrium processes. However, TIR and AIR have consistent outcomes in terms of signal processing~\cite{Yao2020CNS}. Considering double-value relationships, TIR and AIR are equivalent in measuring the probabilistic 
difference between ups and downs, and equal values imply time-reversibility and amplitude-reversibility at the same time. Even if we consider triple- and multiple-value relationships, TIR and AIR again produce highly consistent results, despite focusing on different joint probabilities. From the statistical definitions, time-reversibility and amplitude-reversibility also have a 
high degree of similarity, and they both characterize nonequilibrium features from the perspective of fluctuation theorems. According to the Evans--Searles fluctuation theorem~\cite{Evans1994,Evans2002,Searles2004}, the dissipation function $\Omega_{t}$ takes on two arbitrary values $A$ and $-A$, and their relative probabilities describes the asymmetry of $\Omega_{t}$ in the 
distributions over a particular ensemble of trajectories. If we consider trajectories of duration $t$ by selecting the initial coordinates $\Gamma_{0}$, then $\Omega_{t}$ takes on some values between $A$ and $A \pm dA$, and the probability density can be obtained as
\begin{eqnarray}
	\label{eq0_}
	p(\Omega_{t}=A)=\int d \Gamma_{0}\delta[\Omega_{t}(\Gamma_{0})-A]f(\Gamma_{0},0).
\end{eqnarray}

Note that $\Gamma_{0}$ is merely a dummy variable of integration. The conjugate probability density is

\begin{eqnarray}
	\label{eq0}
	p(\Omega_{t}=-A)=\int d \Gamma_{t}^{*}\delta[\Omega_{t}(\Gamma_{t}^{*})+A]f(\Gamma_{t}^{*},0).
\end{eqnarray}

Given the definitions \cite{Sevick2008,Evans2016} of $\Omega_{t}(\Gamma_{t}^{*})=-\Omega_{t}(\Gamma_{0})$, $\Omega_{t}(\Gamma_{0})=ln [\frac{f(\Gamma_{0},0)}{f(\Gamma_{t}^{*},0)}]-\int_{0}^{t} \Lambda (\Gamma_{s}) ds$, and $|\frac{d\Gamma_{t}}{d\Gamma_{0}}|=\frac{\delta V (\Gamma_{t})}{\delta V (\Gamma_{0})}=exp [\int_{0}^{t} \Lambda (\Gamma_{s}) ds]$, we have 

\begin{eqnarray}
	\label{eq1_}
	p(\Omega_{t}=-A)=exp (-A)\int d \Gamma_{0}\delta[\Omega_{t}(\Gamma_{0})-A]f(\Gamma_{0},0) ,
\end{eqnarray}
based on which we obtain the steady state Evans-Searles fluctuation theorem \cite{Sevick2008,Evans2016,Seife2012}:

\begin{eqnarray}
	\label{eq1}
	\frac{p(\Omega_{t}=A)}{p(\Omega_{t}=-A)}=exp(A).
\end{eqnarray}

Based on the Evans–Searles fluctuation theorem, the time dissipation function $\Omega_{T}$ for arbitrary values $A$ and $-A_{T}$ and the amplitude dissipation function $\Omega_{A}$ for arbitrary values $A$ and $-A_{A}$ should have a high degree of consistency, that is,

\begin{eqnarray}
	\label{eq2}
		\frac{p(\Omega_{T}=A)}{p(\Omega_{T}=-A_{T})} =exp(A_{T}) \approx exp(A_{A})=\frac{p(\Omega_{A}=A)}{p(\Omega_{A}=-A_{A})}.
\end{eqnarray}

Therefore, TIR and AIR both provide a measure whereby fluctuation theorems can characterize nonequilibrium features. TIR and AIR are measured from the perspective of the probability difference between forward and backward sequences. This requires the whole sequence and its time- or amplitude-reversal to be obtained, which is convenient in operational terms, but does not incorporate real-time characteristics. If the sequence is too large, or even if the system produces an uninterrupted process, this forward--backward approach is not applicable for the quantification of TIR or AIR. Alternatively, the probabilistic differences of time- and amplitude-symmetry vectors could be employed. Taking TIR as an example, the probabilistic differences between 
forward--backward series is the same as those of the symmetric vectors~\cite{Yao2020ND}. That is why TIR is also defined as temporal asymmetry~\cite{Costa2005,Costa2008,Guzik2006,Porta2008,Diks1995}. The probabilistic differences between symmetric joint distributions of a process have real-time features that enable TIR to be measured through the generation of signal data. However, in numerical analysis, the correspondence between the symmetry vectors and their simplified forms must be carefully considered. For example, in permutation analysis, original permutation is not a direct reflection of the vector's temporal structure, and so the symmetry original permutations does not always correctly reflect the symmetry vectors~\cite{Yao2022PLA}. Therefore, symmetric original permutations for quantifying TIR are incorrect, but could be used to quantify AIR, while symmetric amplitude permutations could be used as alternatives for quantifying TIR~\cite{Yao2020CNS}.

\subsection{Original and amplitude permutations}
There are two basic ordinal patterns~\cite{Yao2022PLA}: the original permutation (OrP) consists of the indexes of reorganized values in the original vector~\cite{Yao2020ND,Bandt2002,Pessa2020}, while the amplitude permutation (AmP) comprises the positions of original values in the reordered vector and directly reflects the vector's temporal structure~\cite{Bandt2020,Zanin2021P}. Both types of ordinal patterns have widespread applicability in simplifying time series and extracting their structural dynamics.

Given the series $X(t)=\{x(1),\ldots,x(t),\ldots,x(L)\}$ with length $L$, vectors with length $m$ and delay $\tau$~\cite{Yao2022PLA} are reconstructed as

\begin{eqnarray}
	\label{eq3}
	X_{m}^{\tau}(t)=\{x(t),x(t+\tau),\ldots,x(t+(m-1)\tau)\} .
\end{eqnarray}

Considering the vector $X(i)=\{x(i_{1}),\ldots,x(i_{i}),\ldots,x(i_{m})\}$ in the reconstructed $X_{m}^{\tau}(i)$, we reorder the elements in ascending order as $X(j)=\{x(j_{1}),\ldots,x(j_{j}),\ldots,x(j_{m})\}$ such that $x(j_{1})<\cdots < x(j_{j})<\cdots <x(j_{m})$. Note that $i$ denotes the index of element in the original vector $X(i)$, and $j$ denotes the index of element in the reordered vector $X(j)$.

To construct AmP~\cite{Bandt2020,Zanin2021P,Yao2019E}, we rewrite the index of $X(i)$ according to $X(j)$ as follows:

\begin{eqnarray}
	\label{eq4}
	X(i,j)=\{x(i,1),x(i,2),\ldots,x(i,j-1),x(i,j),x(i,j+1),\ldots,x(i,m-1),x(i,m)\} ,
\end{eqnarray}
where $j$ increases from 1 to $m$, and $i$ is the position of original value $x(i)$ in the reordered vector $X(j)$. The series of $i$ is AmP, $\pi_{i}=(i_{1},i_{2},\ldots,i_{j-1},i_{j},i_{j+1}, \ldots, i_{m-1},i_{m})$.

Corresponding to the AmP, we rewrite the index of $X(j)$ according to $X(i)$ to construct the OrP~\cite{Bandt2002} as follows:
\begin{eqnarray}
	\label{eq5}
	X(i,j)=\{x(1,j),x(2,j),\ldots,x(i-1,j),x(i,j),x(i+1,j),\ldots,x(m-1,j),x(m,j)\} ,
\end{eqnarray}
where $i$ increases from 1 to $m$, and $j$ is the position of the reordered $x(j)$ in the original vector $X(i)$. The series consisting of $j$ is OrP, $\pi_{j}=(j_{1},j_{2},\ldots,j_{i-1},j_{i},j_{i+1}, \ldots, j_{m-1},j_{m})$.

Equal values play important roles in the construction of permutations and in permutation analysis~\cite{Yao2019E,Bian2012,Zunino2017,Yao2020APL}. In some real-world signals, equal values are widely distributed and convey important information about system conditions \cite{Yao2019E,Yao2021DES}. The existence of equal values affects the construction of ordinal patterns and can significantly change the probability distributions of permutations. More importantly, equal values could generate self-symmetric vectors (i.e.,~the symmetric vector is itself), which have significant physical implication, i.e.,~time reversible or temporal symmetry~\cite{Yao2020CNS,Yao2020ND,Yao2019E}. Therefore, in permutation analysis, equal values should not be numerically broken by adding small random perturbations or be arranged simply according to their occurrence orders~\cite{Bandt2002,Bian2012}. To deal with equal values, we could modify their indexes to be the same as, for example, the smallest or largest index in each group of equalities~\cite{Yao2022PLA}. First, we arrange the equal values in neighboring orders according to their orders of occurrence as

\begin{eqnarray}
	\label{eq5_}
	\cdots<x(i_{1},j_{1})=x(i_{2},j_{2})< \cdots <x(i_{3},j_{3})=x(i_{4},j_{4})=x(i_{5},j_{5})<\cdots.
\end{eqnarray}

We rewrite the indexes of equal values to be the same as those in their corresponding groups~\cite{Yao2022PLA}, e.g.,~the smallest index as $\cdots<x(i_{1},j_{1})=x(i_{1},j_{1})< \cdots <x(i_{3},j_{3})=x(i_{3},j_{3})=x(i_{3},j_{3})<\cdots$, as suggested by Bian et al.~\cite{Bian2012}, or the largest index as $\cdots<x(i_{2},j_{2})=x(i_{2},j_{2})< \cdots <x(i_{5},j_{5})=x(i_{5},j_{5})=x(i_{5},j_{5})<\cdots$, and then modify the corresponding values in permutation. OrP and AmP reflect the structures of the reordered and original vectors, respectively; therefore, the equal-value indexes in OrP are consecutive, whereas those in AmP might not be.

\begin{figure}[htb]
	\centering
	\includegraphics[width=15.5cm,height=5cm]{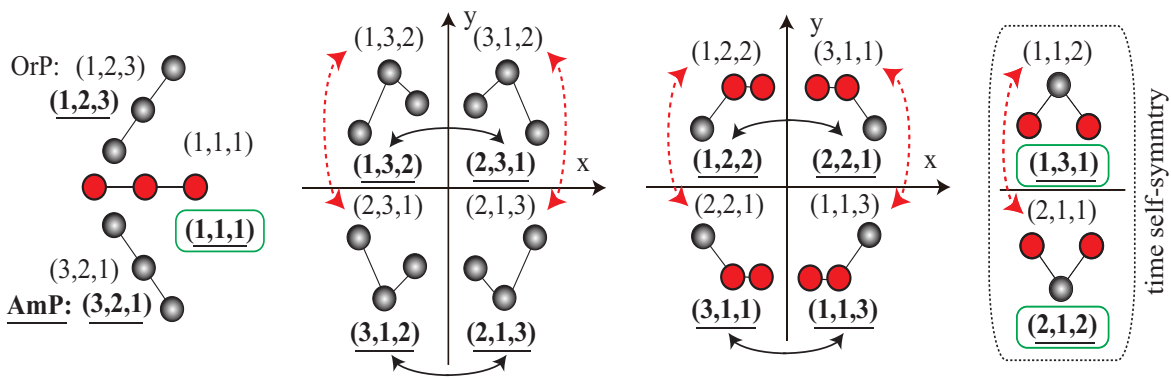}
	\caption{OrP and AmP of triple-value vectors. Equal values are shown in red, and their indexes are modified to be the smallest ones in the equal-values group. Solid black arrows point to symmetric AmPs (bold and underlined), and their vectors are time- and $y$-axis symmetric. Dashed red arrows point to symmetric OrPs, and their vectors are amplitude- and $x$-axis symmetric. AmPs (in green boxes) of temporally self-symmetric vectors are all self-symmetric.}
	\label{fig2}
\end{figure}

Taking triple-value vectors as an example, the association of vector structures and their OrPs and AmPs are illustrated in  Fig.~\ref{fig2}.

Given conventional wisdom concerning time symmetry, i.e.,~$y$-axis symmetry, AmP directly reflects the vector's temporal structure, while OrP is a indirect alternative~\cite{Yao2022PLA}. AmPs of symmetric vectors are symmetric, and those of temporally self-symmetric vectors are all symmetric, as illustrated by Fig.~\ref{fig2}. However, considering amplitude symmetry, i.e.,~the x-axis symmetry, OrPs of symmetric vectors, linked by those dashed red arrows, are also symmetric. The differences between OrP and AmP are irrelevant if the permutation is used as a symbol, or a label, of the vector, such as in permutation entropy~\cite{Bandt2002,Bandt2020}. In contrast, if ordinal pattern is detected as a reflection of the vector structure, OrP and AmP should be selected accordingly to avoid possibly misleading or erroneous analysis, particularly in the quantification of TIR and AIR~\cite{Yao2020CNS,Yao2020ND,Yao2019Sym}.  

Further, if we require real-time characteristic and consider time- and amplitude-symmetric vectors for quantifying TIR and AIR, we must pay attention to the construction of OrP and AmP and their relationships with the vector structures~\cite{Yao2022PLA}. We should measure the probabilistic differences of symmetric AmPs to quantify TIR, because AmP directly reflects the temporal structure of vectors. Similarly, we should calculate the probabilistic differences of the symmetric OrPs to measure AIR. Otherwise, if we construct time- or amplitude-reverse series and employ the forward--backward method to quantify TIR or AIR, the two basic ordinal patterns will be essentially the same because we are simply calculating the probabilistic differences between pairs of same permutations in the original and reverse processes.

\subsection{Joint permutation for TIR and AIR}
After extracting the local dynamical structures, note that neither the individual OrP nor AmP includes global features of the vectors in series. Let us extract global information of the vectors to improve the ordinal pattern.

Given the series $X(t)=\{x(1),\ldots,x(t),\ldots,x(L)\}$, we subtract its mean $X(t)=X(t)-\mu$, generate its time- and amplitude-reversal series $X(-t)$ and $-X(t)$, and construct their space vectors $X_{m}^{\tau}(t)$, $X_{m}^{\tau}(-t)$, and $-X_{m}^{\tau}(t)$ according to Eq.~(\ref{eq3}). Taking original $X(t)$ as an example, the vector $X(i)$ in the reconstructed space $X_{m}^{\tau}(t)$ is first simplified using its AmP $\pi_{i}$. We then detect the global information $s_{i}$ of the vector as follows:

\begin{eqnarray}
	\label{eq6}
	s_{i}=
	\left\{
	\begin{array}{lr}
		1,  \mu_{i} > \mu \\
		0, \mu_{i} \leq \mu,
	\end{array}
	\right.
\end{eqnarray}
where $\mu_{i}$ is the mean of vector, and the mean of series $\mu=1/L \sum x(t)$ is zero. Joint permutation of vector $X(i)$ is a combination of the global $s_{i}$ and $\pi_{i}$ of the form $c_{i}=\{ s_{i}; \pi_{i} \}$.

\begin{figure}[htb]
	\centering
	\includegraphics[width=8.3cm,height=3cm]{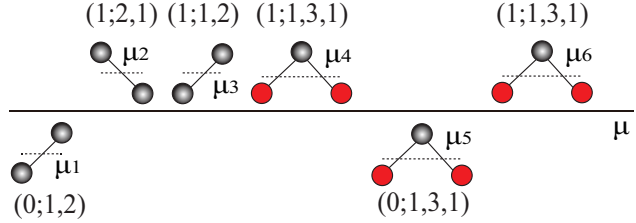}
	\caption{Vectors and their joint permutations. The AmP $\pi_{i}$ is applied to the joint permutation $c_{i}=\{ s_{i}, \pi_{i} \}$. $\mu$ is the mean of the series, $\mu_{i}$ denotes the mean of each vector, and their relationships determine the global $s_{i}$. Red elements in the vectors represent equal values and their indexes are modified to be the smallest value of `1' in the equal-values group.}
	\label{fig3}
\end{figure}

Joint permutation refines the ordinal structures. As shown in Fig.~\ref{fig3}, the ups have permutation (1,2) and are further classified into those below (0;1,2) and above (1;1,2) the mean series. In individual double-permutation analysis, there are three motifs, i.e.,~up, down, and equality, where equality implies time- and amplitude-reversibility at the same time. Therefore, TIR and AIR are the same in the double-value individual permutation~\cite{Yao2020CNS}, while in joint permutation analysis, they are distinguishable. Self-symmetric vectors with the permutation (1,3,1) imply time-reversibility, whereas (0;1,3,1) and (1;1,3,1) differ for the quantification of TIR in the joint permutation scheme.

Through joint ordinal scheme, the original series $X(t)$ is transformed into $C(i)=\{c_{1},c_{2},\ldots,c_{i},\ldots,c_{(L-(m-1)\tau)}\}$, and the probability distributions for the joint permutation sequences are $P(i)=\{p_{1},p_{2},\ldots,p_{i},\ldots\}$. Corresponding to $P(i)$, the time- and amplitude-reversal series $X(-t)$ and $-X(t)$ give $P(-i)=\{p_{-1},p_{-2},\ldots,p_{-i},\ldots\}$ and $P(\mbox{}_{-}i)=\{p_{\mbox{}_{-}1},p_{\mbox{}_{-}2},\ldots,p_{\mbox{}_{-}i},\ldots\}$ for the permutation sequences. The quantification of TIR and AIR could be simplified by the probabilistic differences between the original $P(i)$ and time-reverse $P(-i)$ and between the original $P(i)$ and amplitude-reverse $P(\mbox{}_{-}i)$.

The existence of forbidden permutations (i.e.,~permutations that do not exist for a process) means that there may be permutations that do not simultaneously exist in the forward and backward series. The probability of forbidden permutations is zero, and their corresponding permutations in time- or amplitude-reversal series are single permutations \cite{Yao2020ND,Yao2019Sym}. If we consider the global information of vectors, more forbidden joint permutations will be generated. The probabilistic difference between a forbidden permutation and its corresponding single permutation should be zero or mathematically infinite, which is a disadvantage to quantitative TIR and AIR~\cite{Yao2020CNS,Yao2020ND,Yao2019Sym}. Therefore, division-based parameters, e.g.,~the Kullback--Leibler difference, might be inappropriate, and subtraction-based indexes, such as $\chi^{2}$~\cite{Daw2003} and $Y_{s}$~\cite{Yao2020CNS,Yao2020ND,Yao2019E,Yao2019Sym}, are expected to be more reliable choices. In this paper, we employ $Y_{s}$ in the form 

\begin{eqnarray}
	\label{eq7}
	Y_{s} \langle p_{i},p_{j} \rangle = p_{i} \frac{p_{i}-p_{j}}{p_{i}+p_{j}},
\end{eqnarray}
where $p_{i} \geq p_{j}$ for calculating the joint permutation probabilistic differences.

The subtraction-based index $Y_{s}$ satisfies following basic characteristics for TIR and AIR~\cite{Yao2020ND}: (1) if $p_{i}=p_{j}$, $Y_{s}=$ 0; (2) if $p_{j}=$ 0, $Y_{s}=p_{i}$; (3) if $p_{i1}-p_{j1}=p_{i2}-p_{j2}$, $Y_{s}$ can be adjusted by $p_{i1}/(p_{i1}+p_{j1})$ and $p_{i2}/(p_{i2}+p_{j2})$. The second of these characteristics enables reliable quantification of TIR and AIR in cases where forbidden joint permutations exist~\cite{Yao2019Sym}. The third feature allows us to discriminate pairs of probability distributions with the same absolute difference.

TIR and AIR based on the probabilistic differences between joint permutations can be quantified by $Y_{s}$ as follows:

\begin{eqnarray}
	\label{eq8}
	\left\{
	\begin{array}{lr}
		TIR= \sum_{i} Ys \langle p_{i}, p_{-i} \rangle = \sum_{i} p_{i} \frac{p_{i}-p_{-i}}{p_{i}+p_{-i}}\\
		AIR= \sum_{i} Ys \langle p_{i}, p_{_{-}i} \rangle = \sum_{i} p_{i} \frac{p_{i}-p_{_{-}i}}{p_{i}+p_{_{-}i}},
	\end{array}
	\right.
\end{eqnarray}
where $p_{i}$ and $p_{-i}$ denote the probability distributions of the joint permutation in the original and time-reverse series, and $p_{i}$ and $p_{_{-}i}$ represent those in the original and amplitude-reverse series.

Because we wish to extract the global information by comparing the means of the series and vectors, the real-time characteristics of symmetry vectors may not be available. In following sections, we focus on the probabilistic differences between joint permutations of forward--backward processes, and we employ AmP as a direct alternative~\cite{Yao2022PLA} to the vector for constructing the joint permutation.

\section{Results}
This section first describes the application of comparative analysis to TIR and AIR using model series according to surrogate theory. The joint permutation TIR and AIR are then applied to real-world epileptic electroencephalography (EEG) signals for nonequilibrium feature detection.

\subsection{Model series test}
\begin{figure}[htb]
	\centering
	\includegraphics[width=16cm,height=13cm]{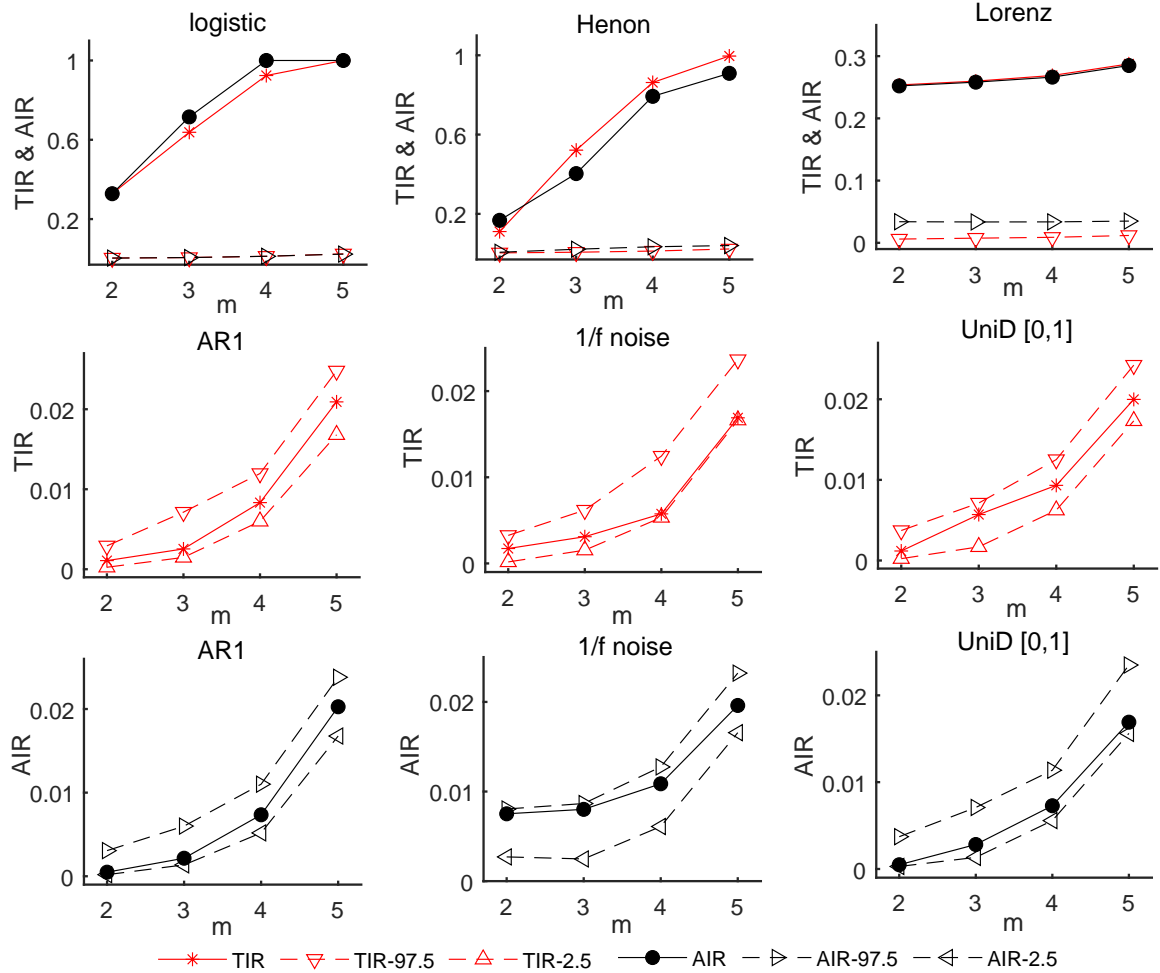}
	\caption{TIR and AIR of chaotic and stochastic series. For logistic equation, $r=$ 4 and $x_{1}=$ 0.01; for Henon and  Lorenz systems, $\alpha=$ 1.4, $\beta=$ 0.3, $\sigma=$ 10,  $b=$ 8/3, and $r=$ 28. Initial values for Henon series are $x_{1}=y_{1}=$ 0.01, and those for Lorenz system are  $x_{1}=y_{1}=$ 0, $z_{1}=1*10^{-10}$. The x-components of Henon and Lorenz systems are used. `AR1' and `UniD~[0,1]' represent the first-order  autoregressive series and uniform distribution in~[0,1],  respectively. Data length of the model series is 50400. The 97.5th and 2.5th percentiles of the TIR and AIR of surrogate data are denoted as `TIR-97.5', `TIR-2.5', `AIR-97.5', and `AIR-2.5'.}
	\label{fig4}
\end{figure}

Let us generate model series and their surrogate series. Logistic equation $x_{t+1}=r \cdot x_{t} (1- x_{t})$, Henon map $x_{t+1}=1-\alpha \cdot x^{2}_{t}+y_{t}$, $y_{t+1}=\beta \cdot x_{t}$, and Lorenz system $dx/dt=\sigma (y-x)$, $dy/dt=x(r-z)-y$, $dz/dt=xy-bz$ are used to construct chaotic model sequences. First-order autoregressive process ($x_{t+1}=\delta x_{t} + \xi_{t}$, where $\xi_{t}$ is a Gaussian series with zero mean and unit variance and $\delta=$ 0.3), 1/f noise, and a uniform distribution in~[0,1] are then employed to generate stochastic series. We use the improved amplitude-adjusted Fourier transform (iAAFT)~\cite{Schrei1996,Schreiber2000} to produce 500 sets of surrogate data for each model series. Surrogate data produced by the iAAFT iteration scheme have same autocorrelations and probability distributions as the given data, and their spectra are practically indistinguishable~\cite{Schrei1996}. Joint permutation TIR and AIR of the chaotic and stochastic model series and their surrogate data are illustrated in Fig.~\ref{fig4}.

According to Fig.~\ref{fig4}, joint permutation TIR and AIR are effective according to the surrogate theory. TIR and AIR of the logistic, Henon, and Lorenz model series are consistently greater than the 97.5th percentiles of their surrogate datasets. In contrast, TIR and AIR values of the stochastic AR1, 1/f noise, and uniform distribution~[0,1] series are all between the 2.5th and 97.5th percentiles of their surrogates.

\begin{figure}[htb]
	\centering
	\includegraphics[width=16cm,height=8cm]{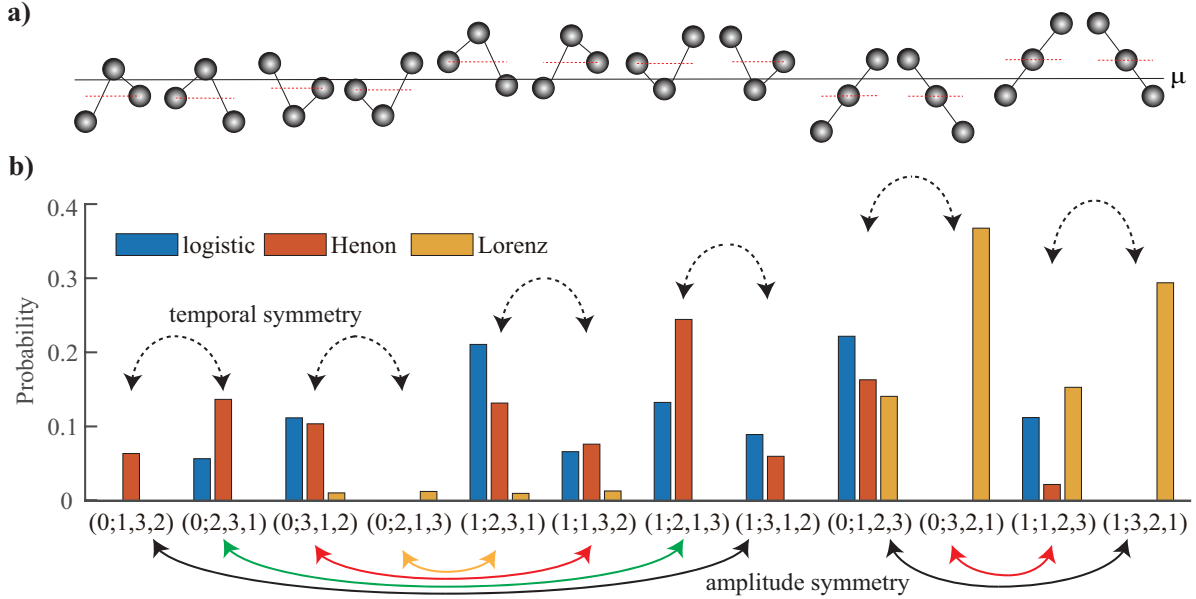}
	\caption{Probability distributions of joint permutations of chaotic logistic, Henon, and Lorenz model series when $m=$ 3. AmPs of the vectors are employed. (a) Twelve simplified vectors of the chaotic series that have no equal values. $\mu$ denotes the mean series, and the red dashed line in each vector represents the vector's mean. (b) Histogram of probability distributions of joint permutations in chaotic series corresponding to the vectors in (a). Joint amplitude permutations of vectors in (a) are shown under the histogram in (b). Temporal-symmetric vectors and their joint permutations for quantifying TIR are linked by dashed double-arrows above the histogram, and amplitude-symmetric vectors and their joint permutations for quantifying AIR are connected by solid double-arrows below the plot.}
	\label{fig5}
\end{figure}

Note that the probabilistic differences of joint permutations of forward--backward series are equivalent to those of symmetric vectors' joint permutations in terms of TIR and AIR. Thus, we compare the pairs of joint permutations of temporal- and amplitude-symmetric vectors to further analyze the TIR and AIR. Taking the three chaotic model series as examples, let us demonstrate the probability distributions of joint permutations comparatively to quantify the TIR and AIR. For the chaotic logistic, Henon, and Lorenz series, the distribution of equal states (DES)~\cite{Yao2021DES} is given by
\begin{eqnarray}
	\label{eq9}
	DES=\frac{N(x_i=x_{i+\tau})}{L-\tau},
\end{eqnarray}
where $N(x_i=x_{i+\tau})$ is the number of neighboring equal values $x_i=x_{i+\tau}$ that are both zero when $\tau=$ 1 and 2. The three chaotic series do not have neighboring equal values, thus having no equal-value permutation when $m=$ 3. A histogram of the probability distributions of joint permutations when $m=$ 3 for these chaotic model series is illustrated in Fig.~\ref{fig5}.

Comparing the distributions of joint permutations for these chaotic model series, we find that the TIR and AIR permutations give highly consistent results in Fig.~\ref{fig4}, but are shown to be different in Fig.~\ref{fig5}. In individual permutation analysis, TIR and AIR share some pairs of vectors, e.g.,~the symmetric all-up and all-down vectors. If only double-value vectors are considered, TIR and AIR measuring the up--down differences are the same~\cite{Yao2020CNS}. In particular, all-equal vectors indicate both time- and amplitude-reversibility. In joint permutation irreversibility, the vector's global information refines the ordinal patterns. TIR calculates the pairs of vectors with same global symbol, while AIR targets those with different global symbols. As shown in Fig.~\ref{fig5}, TIR and AIR do not share same pairs of joint permutations, and so they quantify the fluctuation theorem of joint permutations differently in terms of nonequilibrium characteristics.

According to the TIR and AIR of chaotic series in Fig.~\ref{fig4} and the probability distributions of joint permutations in  Fig.~\ref{fig5}, TIR and AIR have a high degree of numerical similarity. Taking Lorenz series as an example, the joint permutation pairs (0;1,3,2), (0;2,3,1) and (1;2,1,3), (1;3,1,2) are time-symmetric vectors that are forbidden permutations, and their contributions to quantifying TIR are zero. At the same time, the pairs (0;1,3,2), (1;3,1,2) and (0;2,3,1), (1;2,1,3) are joint permutations of amplitude-symmetric vectors, and their contributions to quantifying AIR are also zero. The probabilistic differences in the other four pairs of joint permutations of time-symmetric vectors are 0.00105, 0.00186, 0.16412, and 0.09278, and those of amplitude-symmetric vectors are 0.00144, 0.00145, 0.10364, and 0.15165, respectively. Lorenz series has TIR and AIR values of 0.25983 and 0.25820 when $m=$ 3, respectively, which almost overlap, as shown by Fig.~\ref{fig4}. Numerical simulations suggest that the probabilistic differences of corresponding TIR and AIR joint permutations for logistic and Henon series also have a high degree of similarity.

From the model series analysis, we learn that TIR and AIR measure different nonequilibrium characteristics, that is, they target different pairs of joint permutations for the calculation of probabilistic differences. At the same time, TIR and AIR both measure the fluctuations in joint permutation distributions and they give highly similar numerical results in model data analysis.

\subsection{TIR and AIR in epileptic EEGs}
Human brain is a highly complex system manifesting nonequilibrium features and subjected to various external environmental factors and internal physiological or pathological conditions~\cite{Fang2019,Lynn2021,Stefanov2020}. Epilepsy is a common chronic neurological disease, characterized by recurrent seizures, which affects people of all ages worldwide and may lead to permanent brain damage~\cite{epilepsy,Moshe2015,Lehner2014}. Many underlying disease mechanisms can lead to epilepsy, and at onset of a seizure, a group of brain cells generates excessive electrical discharges that manifest as interictal spikes in brain electrographic signals. These spikes have been reported to have nonequilibrium characteristics~\cite{Martin2018,Yao2020ND,Donges2013,Schindle2016}. In this section, we use the joint permutation TIR and AIR to characterize real-world epileptic EEGs obtained from the public Bonn database~\cite{Andrze2001}.

\begin{figure}[htb]
	\centering
	\includegraphics[width=7.5cm,height=5.5cm]{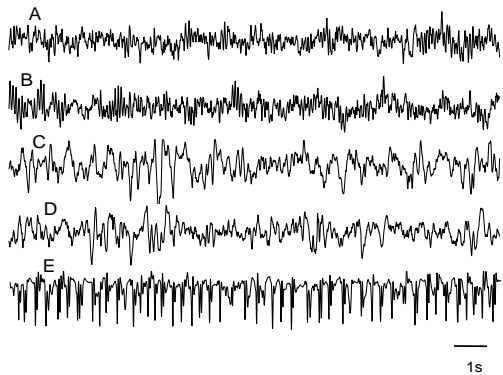}
	\caption{Example EEG series from five sets of Bonn epileptic data. Amplitude distributions of EEGs A–D are centered around approximately 100 $\mu$V, and that of EEG E is centered around 1000 $\mu$V.}
	\label{fig6}
\end{figure}

Bonn epileptic database consists of five datasets (A--E) in which the EEGs contain 100 single-channel segments with a duration of 23.6 s. Sets A (eyes open) and B (eyes closed) contain surface EEG recordings from five healthy subjects, obtained according to the international 10--20 system. Sets C (from the hippocampal formation of the opposite hemisphere of the brain) and D (from within the epileptogenic zone) consist of EEGs during seizure-free intervals. Segments of set E are collected from the intracranial electrodes and contain seizure activity. All EEG segments are recorded with a sampling frequency of 173.61~Hz, a bandpass filter of 0.53--40~Hz, and 12-bit analog-to-digital conversion. Electrodes that contain strong eye movement artifacts in sets A and B and pathological activities in sets C, D, and E are omitted. Detailed information about Bonn epileptic EEGs can be found in Ref.~\cite{Andrze2001}. EEGs from the same group of subjects are related, and those from different groups of subjects are independent. Given the characteristics of each group of EEGs and their relationships, we employed the nonparametric Mann--Whitney U-test for independent samples, i.e.,~EEGs from the healthy and epileptic subjects, and the Wilcoxon test to measure the statistical differences among related samples, i.e.,~sets A and B as well as sets C, D, and E. Example time series of Bonn EEGs are shown in Fig.~\ref{fig6}.

\begin{figure}[htb]
	\centering
	\includegraphics[width=16cm,height=10cm]{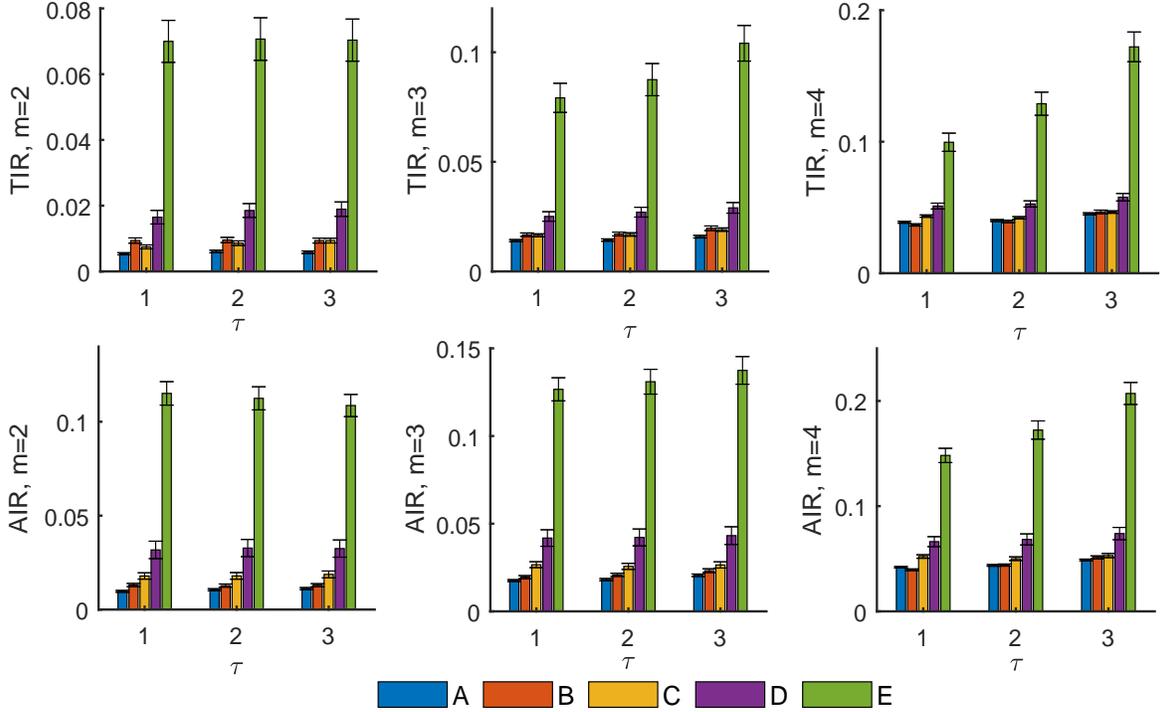}
	\caption{TIR and AIR (mean $\pm$ standard error) of the Bonn epileptic data sets.}
	\label{fig7}
\end{figure}

According to our previous report~\cite{Yao2021DES}, ictal EEGs in set E manifest abnormally low distribution of equal values because of neural firing, whereas seizure-free EEGs in sets C and D have larger DES than the healthy ones in sets A and B. Because equal values contain important system information and are important for the construction of ordinal patterns, equal-value permutation is further proved to be necessary in the quantitative TIR and AIR of epileptic EEGs. The TIR and AIR results of the five groups of EEGs are shown in Fig.~\ref{fig7}. Bonn epileptic EEGs share consistent TIR and AIR characteristics. Seizure signals in set E exhibit the greatest degree of irreversibility, and seizure-free signals of sets C and D have higher nonequilibrium characteristics than the healthy brain electric signals in sets A and B.

The nonequilibrium characteristics of epileptic seizure EEGs in set E, as quantified by both TIR and AIR, are abnormally higher than those of the EEGs obtained under other conditions. The significantly high TIR and AIR of seizure EEGs are in line with epileptic pathological brain states. Pathologically, epilepsy manifests as recurrent epileptic seizures caused by sudden development of synchronous neuronal firing in cerebral cortex~\cite{Lehner2014,Schindle2016,Andrze2001,Iasemi2003,Cui2019}. During the onset of a seizure, a group of brain cells exhibits ictal activity, i.e.,~abnormally excessive electrical discharges. Brain's electrical activity during epileptic seizures has been widely reported to be extremely complex, e.g.,~amplitude fluctuations~\cite{Yao2021DES}, aberrant connectivity~\cite{Cui2019}, and nonlinear dynamics~\cite{Yao2020ND,Donges2013,Andrze2001,Kulp2017}. Therefore, joint permutation TIR and AIR correctly and effectively detects the significantly abnormal nonequilibrium features that reflect epileptic seizure activities.

After onset of an epileptic discharge, partial seizures might remain localized~\cite{Iasemi2003,Cui2019}. Although exact time intervals between the ictal recordings in set E and postictal EEGs in sets C and D are not clear, the characteristics of EEGs in sets C and D indicate the pathological seizure impacts~\cite{Yao2020ND,Yao2021DES,Andrze2001}. Seizure-free interictal signals of EEGs in sets C and D have higher TIR and AIR than those of sets A and B. Additionally, set D has much larger TIR and AIR within epileptogenic zone, suggesting that epilepsy increases the nonequilibrium characteristics of brain's electric activities.

Both joint permutation irreversibility parameters, particularly AIR, effectively characterize pathological features of epileptic EEGs. TIR and AIR provide consistent findings in epileptic EEG analysis, but AIR produces better results according to statistical tests. When $m=$ 2, TIR value of healthy EEGs in set B is greater than for the epileptic signals in set C, which is different from other results. AIR produces consistent results, i.e.,~seizure set E manifests the largest AIR and epileptic seizure-free signals of sets C and D have larger AIR values than healthy signals of sets A and B. According to the overall statistical test results, AIR more effectively identifies the epileptic seizure-free EEGs from healthy EEGs.

Overall, TIR and AIR are fundamentally different in terms of both their statistical concepts and numerical results. However, the two parameters are closely related in that they both measure nonequilibrium characteristics from the perspective of a fluctuation theorem, and they produce consistent results in model series and real-world EEG analysis.

\section{Discussions}
The expressions determining TIR and AIR have been simplified by using joint permutations and comparative analysis of these expressions has been conducted through model series and brain electrical data. This investigation has demonstrated that several issues require further discussion.

The first issue is the implication of the high consistency of TIR and AIR in nonequilibrium analysis. TIR and AIR are fundamentally different descriptors for nonequilibrium processes. TIR measures time-reversal differences, whereas AIR calculates amplitude-reversal differences---these are different theoretical concepts and target different pairs of joint permutations in numerical simulations~\cite{Yao2020CNS}. However, TIR and AIR have similar statistical definitions and produce highly consistent results when they are applied to signal processing. The direct reason lies in the highly similar probability distributions of pairs of time- and amplitude-reversal joint permutations, as the corresponding differences for quantifying TIR and AIR are very similar. The fundamental reason is that both TIR and AIR provide a measure for fluctuation theorem as a means of characterizing nonequilibrium systems. The similarities between TIR and AIR inspire us to go beyond the limitations of either, and elucidate nonequilibrium systems using fluctuation theorem from a broader perspective. For example, traditional probability estimation 
based on space vector distances is not suitable for quantifying TIR, so we could instead measure the up--down fluctuations of space vector distances in response to some tolerance in reconstructed state space~\cite{Yao2021CSF}. Note that the visibility of TIR~\cite{Lacasa2012,Lacasa2015,Flanagan2016}, in terms of measuring the probabilistic differences among in--out degrees, in fact 
describes a form of fluctuation theorem from the perspective of a visibility graph network. During the construction of a visibility graph~\cite{Lacasa2008,Luque2009}, vectors are transformed based on visual connections, thus making visibility TIR different from the strict statistical concept of TIR. Inspired by the consistency of TIR and AIR, as well as their implications, we can also measure the fluctuation of coupled systems with effective connectivity, and detect nonequilibrium characteristics of 
networks by calculating the in--out differences of local individual nodes and of the global network. Therefore, comparative analysis of TIR and AIR contributes to our understanding of nonequilibrium systems in terms of fluctuation theorem, and provides a novel idea for broadening the quantitative measurement of nonequilibrium.

Note that, in quantitative TIR or AIR, the forward--backward probabilistic differences and symmetric vectors probabilistic differences are equivalent~\cite{Yao2020ND}. Symmetric method has desirable real-time performance, while the forward--backward approach is more reliable in applications. Currently, large amounts of data are generated by physiological, meteorological, or engineering systems, and some uninterrupted signal collection is inevitable in the real-time recordings. Under these conditions, 
obtaining and reversing all of the data are impractical, so the probabilistic differences of time- and amplitude-symmetric vectors for quantitative TIR and AIR are preferable. On other hand, the symmetric approach has certain limitations if the vector is not reliably associated with its alternative. For example, original permutation does not directly reflect the temporal structure of a vector, and so the symmetric original permutations used as alternatives to the temporal symmetry vectors for quantitative TIR are incorrect and will lead to conceptual errors in numerical simulations~\cite{Yao2019Sym}. This is why we 
proposed the concept of amplitude reversibility~\cite{Yao2020CNS}. In addition, joint permutations for symmetric vectors require more attention. For both the original form and amplitude permutation, temporal symmetric vectors for quantitative TIR contain same global information, whereas amplitude symmetric vectors for AIR have different global symbols, as shown in Fig.~\ref{fig5}a. If we employ forward--backward method, we need only calculate the probabilistic differences of same permutations between the original-reverse series, which is more convenient and reliable. Therefore, if conditions permit, the forward--backward approach should be used for quantitative TIR and AIR because of its higher reliability in applications. If the symmetric method is adopted for its real-time performance, particular attention should be paid to the associations of the vector and its alternative to avoid possible mistakes and conceptual errors.

Next, let us focus on the significance of equal values in permutation analysis. In logistic, Henon, and Lorenz chaotic series, the distributions of neighboring equal values are all zeros, and equal values are rare in epileptic EEGs~\cite{Yao2021DES}. However, equal values play an important role in permutation analysis~\cite{Bian2012}. As shown by Fig.~\ref{fig2}, equalities are 
necessary for characterizing reliable vector temporal structures. Without equalities, there would be 6 (3!) kinds of order patterns; in fact, there are 7 more representations with equal values, reflecting the comprehensive relationships among vectors and the OrP/AmP. Equal values are neglected because they are assumed to be very rare and do not contain information about the 
system~\cite{Bandt2002}, although this assumption is also proved to be incorrect. Diseased heart rates with reduced variability have almost 46\% neighboring equal heartbeats~\cite{Yao2019E}, and equal values significantly affect the probabilities of permutations and even generate totally contradictory findings~\cite{Yao2020APL}. More importantly, the decreased distribution of equal heartbeats is in line with reduced variability in heart rates, and might serve as a clinical indicator to sudden death in 
patients recovering from myocardial infarctions. In EEGs, the distribution of equal states is inherently rooted in analog-to-digital conversion~\cite{Yao2021DES}, enabling precise quantitative amplitude fluctuations with the characteristics of simplicity, high stationarity, and robustness under different physiological and pathological conditions. Significantly, in permutation irreversibility, equal values might generate self-symmetric vectors, which have special physical meaning with respect to time-reversible or temporal symmetry~\cite{Yao2020CNS,Yao2020ND,Yao2019E}. All-equal vectors imply time- andamplitude-reversibility at the same time. Therefore, equal-values ordinal patterns are a necessity for reliable research in permutation analysis.

Forbidden permutations are reported to be closely associated with system structures~\cite{Kulp2017,David2020,Amigo2008}, but they have negative effects on permutation irreversibility. Note that a forbidden permutation might generate a single permutation in the quantitative TIR and AIR. For example, regarding the logistic series in Fig.~\ref{fig5}, (0;1,3,2), (0;2,1,3), (0;3,2,1), and (1;3,2,1) are all forbidden permutations; therefore, (0;2,3,1), (0;3,1,2), (0;1,2,3), and (1;1,2,3) for TIR and (1;3,1,2), (1;2,3,1), (1;1,2,3), and (0;1,2,3) for AIR are all single permutations. When the length of the vector (i.e.,~the dimension $m$) is greater than or equal to 5, all of the existing joint permutations are single permutations. In this case, we need to calculate the difference between zero and a nonzero probability distribution to quantify TIR and AIR. This situation makes division-based 
parameters unsuitable, because the probabilistic difference between forbidden permutation and its single permutation will be zero or infinity, which is inappropriate for the quantification of TIR and AIR~\cite{Yao2020CNS,Yao2020ND,Yao2019Sym}. During the construction of a visibility graph~\cite{Lacasa2008,Luque2009}, we find that there are also in and out degrees that do not simultaneously exist in the forward and backward time series, analogous to the situation of forbidden permutations. However, 
most researchers have employed division parameters such as the Kullback--Leibler distance. We believe there are two reasons for this: (1) the fluctuation theorems, whether the Evans--Searles form $\frac{P(\Omega=A)}{P(\Omega=-A)}=exp(A)$~\cite{Evans1994,Evans2002,Searles2004,Sevick2008,Evans2016} or the Crooks expression 
$\frac{P_{F}(+w)}{P_{R}(-w)}=exp(+w)$~\cite{Sevick2008,Evans2016,Seife2012,Crooks1999}, have an exponential form, and it is mathematically convenient to measure the fluctuations with division-based measures; (2) division parameters are frequently used in computational physics and have been intensively studied in information theory. However, because certain characteristics of quantitative TIR and AIR are based on coarse-grained series, such as the symbolic time series and visibility graph, there are single vectors similar to the single permutations that lead to forbidden symbols or forbidden visibilities. Thus, subtraction-based parameters are more suitable for quantitative analysis of TIR and AIR.

The length $m$ and delay $\tau$ are crucial parameters for the construction of permutations, and their effects on quantitative TIR and AIR should also be examined. According to the conceptual definitions of TIR and AIR~\cite{Yao2020CNS,Weiss1975,Kelly1979}, a reversible process should have the same joint probability distributions as its reverse for all dimensions and delays. Irreversibility is defined as a statistical change in reversed 
series, but no restriction is imposed on that function~\cite{Zanin2021C}. Therefore, as long as any difference between the original and reverse sequences with any dimension or delay can be found, TIR or AIR of the sequence can be proved. For example, the widely accepted up--down differences in the indexes of Costa, Guzik, and Porta~\cite{Costa2008,Guzik2006,Porta2008,Cammarota2007} are in fact equivalent to a specific case of permutation TIR with $m=$ 2 
and $\tau=$ 1. In phase space theory, the selection of $m$ and $\tau$ is vital to the quality of reconstruction. According to 
Takens' theorem, $m$ should satisfy the sufficient condition $m \geq 2d+1$, where $d$ is the fractal dimension of underlying 
attractor~\cite{Zou2019,Kim1999,Kugium1996}. Additionally, $\tau$ should be large enough to prevent redundancy (i.e.,~the attractor is compressed along the identity line), but small enough to avoid irrelevance (i.e.,~the reconstructed attractor becomes causally disconnected)~\cite{Kim1999}. From the perspective of analog-to-digital conversion, increasing time delay is 
consistent with reducing sampling frequency, i.e.,~the frequency reduces from $f$ to $f/\tau$ given the delay $\tau$~\cite{Yao2021DES}. Therefore, $\tau$ should be restricted to ensure that the sampling frequency satisfies the Nyquist sampling law (i.e.,~twice the bandwidth of the target system). There are several methods for estimating $m$ and $\tau$ independently or simultaneously; however, in real-world applications, 'trial and error' methods can be used~\cite{Kugium1996}. Taking the quantification of TIR as an example, $m$ could be selected from 2--5 and $\tau$ from 1--3 by enumeration, and results could then be analyzed accordingly. 

Another issue that should be discussed concerns the surrogate data ofhe Bonn healthy EEGs. TIR and AIR of the epileptic seizure EEGs in set E are larger than the 97.5th percentiles of their surrogate data, suggesting the manifestation of nonlinearity in seizure EEGs. However, some EEG signals from other groups, particular the healthy sets A and B, are between the 2.5th 
and 97.5th percentiles of their surrogate data, indicating that they are compatible with linear stochastic series according to surrogate theory. These findings reflect our previous report on the permutation TIR of epileptic EEGs and the original research on nonlinear deterministic dynamics of these Bonn data~\cite{Yao2020ND,Donges2013,Andrze2001,Kulp2017}. Andrzejak et al.~\cite{Andrze2001} inferred that the huge number of neurons involved in EEG recordings and the highly complicated brain structure are two causes for the acceptance of null hypothesis in surrogate tests. Moreover, surface EEGs are collected under different skull conductivities and via other intermediate tissue; therefore, the filter process might further blur the 
possible dynamical structures. This assumption could be true, but the high complexity of brain structure should be a inherent reason for the nonlinearity of brain activities. Although surface EEGs present randomness or stochasticity, the complicated brain activities also have multiple time scales and multidimensional characteristics. With changes in the scale factor or reconstruction dimension, different or even contradictory results may appear, and this is an important consideration in biological signals analysis. Vital information about physiological or pathological conditions might be revealed from multidimensional reconstructed state space~\cite{Zou2019,Kim1999,Kugium1996} or through multiscale analysis~\cite{Costa2008,Costa2002}, and this topic requires more research.

\section{Conclusions}
TIR and AIR are associated measures for nonequilibrium processes. In this paper, we have conducted comparative analysis of TIR and AIR from their theoretical concepts, and have demonstrated their relationship through numerical experiments based on the joint permutations of model series and real-world brain electrical data. Our major findings can be summarized as follows:

(i) Conceptually, TIR and AIR are different nonequilibrium descriptors in that they measure the time- and amplitude-reversibility characteristics, respectively. These characteristics could also be measured from time- and amplitude-symmetric vector differences. Symmetric vector differences for quantifying TIR and AIR produce real-time performance, but it is necessary to check the correspondence between the vectors and their simplified forms. In numerical simulations based on individual ordinal patterns of symmetric vectors, TIR should be quantified by symmetric AmPs, while AIR should be measured by symmetric OrPs. Different TIR and AIR outcomes appear in the chaotic logistic, Henon, and Lorenz series, as well as in epileptic EEG analysis. The findings regarding these differences between TIR and AIR improve our understanding of nonequilibrium systems and focus our attention on their simplified quantification to avoid conceptual errors.

(ii) Although TIR and AIR target different aspects of nonequilibrium processes, they both provide a measure for the fluctuation theorem of nonequilibrium processes and produce consistent results when applied to signal processing. TIR and AIR both calculate the original-reversal probabilistic difference, and share similarities in terms of statistical concepts and in the characterization of fluctuation theorems. In permutation numerical simulations, probabilistic differences between pairs of joint permutations of TIR and AIR are similar, leading to highly consistent TIR and AIR results in both the model series and real-world data. This consistency suggests we have broader choices for measuring nonequilibrium processes, because we can quantify the fluctuation theorem from the perspectives of the original series, coupling causality, reconstructed spaces, or effective networks.

(iii) OrP specifies the positions of reordered vector values in the original vector, while AmP comprises the positions of original values in the reordered vector and directly reflects the temporal structure of the vector. Equal-values ordinal scheme is necessary because of its reliable vector structure detection and important physical implication (i.e.,~reversibility) of self-symmetric vectors in TIR and AIR. OrP and AmP rely on the comparison of neighboring values, and so neither contains global features of the vectors. The combination of global information and permutation of the vector extracts more details about the temporal structure of ordinal pattern, thus enabling more precise detection of TIR and AIR in nonequilibrium analysis.

(iv) Regarding the epileptic EEGs, the extremely high TIR and AIR of the seizure EEGs are consistent with abnormal firing discharge during onset of a seizure. The irreversibility of seizure-free EEGs is greater than that of healthy control EEGs. These findings suggest that TIR and AIR both reliably characterize epileptic pathological conditions contained in brain electric signals, among which AIR provides better detection of nonequilibrium feature, particular for seizure-free EEGs.

Overall, the comparative analysis of TIR and AIR improves our understanding of the associations between TIR and AIR in quantifying nonequilibrium characteristics, and broadens the scope of quantitative nonequilibrium methods by measuring fluctuations from different system perspectives. Our findings contribute to the development of permutation symbolic time series analysis and to the understanding of nonequilibrium brain electrical activities under different physiological and pathological conditions.

\section{Acknowledgment}
The project is supported by the Natural Science Foundation of Jiangsu Province (Grant No.BK20220383), Natural Science Research Start-up Foundation of Recruiting Talents of Nanjing University of Posts and Telecommunications (Grant No.NY221142), Shanghai Municipal Science and Technology Major Project (Grant No.2018SHZDZX01), Key Laboratory of Computational Neuroscience and Brain-Inspired Intelligence (LCNBI) and ZJLab.

\nocite{*}

\bibliography{mybibfile}

\begin{thebibliography}{10}
\expandafter\ifx\csname url\endcsname\relax
  \def\url#1{\texttt{#1}}\fi
\expandafter\ifx\csname urlprefix\endcsname\relax\def\urlprefix{URL }\fi
\expandafter\ifx\csname href\endcsname\relax
  \def\href#1#2{#2} \def\path#1{#1}\fi

\bibitem{Yao2020CNS}
W.~P. Yao, J.~Wang, M.~Perc, W.~L. Yao, J.~Dai, D.~Guo, D.~Yao, Time
  irreversibility and amplitude irreversibility measures for nonequilibrium
  processes, Communications in Nonlinear Science and Numerical Simulation 96
  (2021) 105688.

\bibitem{Weiss1975}
G.~Weiss, Time-reversibility of linear stochastic processes, Journal of Applied
  Probability 12~(4) (1975) 831--836.

\bibitem{Ropke2013}
G.~Ropke, Nonequilibrium Statistical Physics, John Wiley \& Sons, 2013.

\bibitem{Costa2005}
M.~D. Costa, A.~L. Goldberger, C.~K. Peng, Broken asymmetry of the human
  heartbeat: loss of time irreversibility in aging and disease, Physical Review
  Letters 95~(19) (2005) 198102.

\bibitem{Fang2019}
X.~Fang, K.~Kruse, T.~Lu, J.~Wang, Nonequilibrium physics in biology, Reviews
  of Modern Physics 91~(4) (2019) 045004.

\bibitem{Lynn2021}
C.~W. Lynn, E.~J. Cornblath, L.~Papadopoulos, M.~A. Bertolero, D.~S. Bassett,
  Broken detailed balance and entropy production in the human brain,
  Proceedings of the National Academy of Sciences 118~(47) (2021).

\bibitem{Wan2018}
K.~Y. Wan, R.~E. Goldstein, Time irreversibility and criticality in the
  motility of a flagellate microorganism, Physical review letters 121~(5)
  (2018) 058103.

\bibitem{Martin2018}
J.~H. Martinez, J.~L. Herrera-Diestra, M.~Chavez, Detection of time
  reversibility in time series by ordinal patterns analysis, Chaos: An
  Interdisciplinary Journal of Nonlinear Science 28~(12) (2018) 123111.

\bibitem{Yao2020ND}
W.~P. Yao, J.~Dai, M.~Perc, J.~Wang, D.~Yao, D.~Guo, Permutation-based time
  irreversibility in epileptic electroencephalograms, Nonlinear Dynamics
  100~(1) (2020) 907--919.

\bibitem{Martin2019}
J.~A. Martin~Gonzalo, I.~Pulido~Valdeolivas, Y.~Wang, T.~Wang,
  G.~Chiclana~Actis, M.~d.~C. Algarra~Lucas, I.~Palmí~Cortés,
  J.~Fernández~Travieso, M.~D. Torrecillas~Narváez, A.~A. Miralles~Martinez,
  E.~Rausel, D.~Gómez~Andrés, M.~Zanin, Permutation entropy and
  irreversibility in gait kinematic time series from patients with mild
  cognitive decline and early alzheimer’s dementia, Entropy 21~(9) (2019)
  868.

\bibitem{Zanin2021C}
M.~Zanin, Assessing time series irreversibility through micro-scale trends,
  Chaos: An Interdisciplinary Journal of Nonlinear Science 31~(10) (2021)
  103118.

\bibitem{Rojo2018}
J.~L. Rojo-Alvarez, M.~Martinez-Ramon, J.~Munoz-Mari, G.~Camps-Valls, Digital
  signal processing with Kernel methods, John Wiley \& Sons, 2018.

\bibitem{Marina2007}
D.~Marinazzo, M.~Pellicoro, S.~Stramaglia, Kernel method for nonlinear granger
  causality, Physical review letters 100~(14) (2008) 144103.

\bibitem{Schreiber2000T}
T.~Schreiber, Measuring information transfer, Physical Review Letters 85~(2)
  (2000) 461--464.

\bibitem{Xiong2017}
W.~Xiong, L.~Faes, P.~C. Ivanov, Entropy measures, entropy estimators, and
  their performance in quantifying complex dynamics: Effects of artifacts,
  nonstationarity, and long-range correlations, Physical Review E 95~(6) (2017)
  062114.

\bibitem{Hlavac2007}
K.~Hlavackova~Schindler, M.~Palus, M.~Vejmelka, J.~Bhattacharya, Causality
  detection based on information-theoretic approaches in time series analysis,
  Physics Reports 441~(1) (2007) 1--46.

\bibitem{Costa2008}
M.~D. Costa, C.~K. Peng, A.~L. Goldberger, Multiscale analysis of heart rate
  dynamics: Entropy and time irreversibility measures, Cardiovascular
  Engineering 8~(2) (2008) 88--93.

\bibitem{Guzik2006}
P.~Guzik, J.~Piskorski, T.~Krauze, A.~Wykretowicz, H.~Wysocki, Heart rate
  asymmetry by poincaré plots of rr intervals, Biomedizinische Technik
  Biomedical Engineering 51~(4) (2006) 272--275.

\bibitem{Porta2008}
A.~Porta, K.~R. Casali, A.~G. Casali, T.~Gnecchi-Ruscone, E.~Tobaldini,
  N.~Montano, S.~Lange, D.~Geue, D.~Cysarz, P.~Van~Leeuwen, Temporal
  asymmetries of short-term heart period variability are linked to autonomic
  regulation, American Journal of Physiology Regulatory Integrative \&
  Comparative Physiology 295~(2) (2008) 550--557.

\bibitem{Cammarota2007}
C.~Cammarota, E.~Rogora, Time reversal, symbolic series and irreversibility of
  human heartbeat, Chaos, Solitons \& Fractals 32~(5) (2007) 1649--1654.

\bibitem{Daw2003}
C.~S. Daw, C.~E.~A. Finney, E.~R. Tracy, A review of symbolic analysis of
  experimental data, Review of Scientific Instruments 74~(2) (2003) 915--930.

\bibitem{Meyer2021}
P.~G. Meyer, H.~Kantz, Time reversal symmetry and the difference between
  relaxations and building-up periods, Physical Review E 104~(2) (2021) 024208.

\bibitem{Raul2021}
R.~Salgado-Garcia, C.~Maldonado, Estimating entropy rate from censored symbolic
  time series: A test for time-irreversibility, Chaos: An Interdisciplinary
  Journal of Nonlinear Science 31~(1) (2021) 013131.

\bibitem{Lacasa2008}
L.~Lacasa, B.~Luque, F.~Ballesteros, L.~Jordi, J.~C. Nuno, From time series to
  complex networks: the visibility graph, Proceedings of the National Academy
  of Sciences of the United States of America 105~(13) (2008) 4972--5.

\bibitem{Luque2009}
B.~Luque, L.~Lacasa, F.~Ballesteros, L.~Jordi, Horizontal visibility graphs:
  exact results for random time series, Physical Review E 80~(2) (2009) 046103.

\bibitem{Lacasa2012}
L.~Lacasa, A.~Nunez, E.~Roldan, J.~M.~R. Parrondo, B.~Luque, Time series
  irreversibility: a visibility graph approach, European Physical Journal B
  85~(6) (2012) 217.

\bibitem{Lacasa2015}
L.~Lacasa, R.~Flanagan, Time reversibility from visibility graphs of
  nonstationary processes, Physical Review E 92~(2) (2015) 022817.

\bibitem{Flanagan2016}
R.~Flanagan, L.~Lacasa, Irreversibility of financial time series: a
  graph-theoretical approach, Physics Letters A 380~(20) (2016) 1689--1697.

\bibitem{Donges2013}
J.~F. Donges, R.~V. Donner, J.~Kurths, Testing time series irreversibility
  using complex network methods, Europhysics Letters 102~(1) (2013) 381--392.

\bibitem{Bandt2002}
C.~Bandt, B.~Pompe, Permutation entropy: a natural complexity measure for time
  series, Physical Review Letters 88~(17) (2002) 174102.

\bibitem{Bandt2020}
C.~Bandt, Order patterns, their variation and change points in financial time
  series and brownian motion, Statistical Papers 61~(5) (2020) 1565--1588.

\bibitem{Zanin2021P}
M.~Zanin, F.~Olivares, Ordinal patterns-based methodologies for distinguishing
  chaos from noise in discrete time series, Communications Physics 4~(1) (2021)
  1--14.

\bibitem{Yao2019E}
W.~Yao, W.~Yao, J.~Wang, Equal heartbeat intervals and their effects on the
  nonlinearity of permutation-based time irreversibility in heart rate, Physics
  Letters A 383~(15) (2019) 1764--1771.

\bibitem{Bian2012}
C.~Bian, C.~Qin, Q.~D. Ma, Q.~Shen, Modified permutation-entropy analysis of
  heartbeat dynamics, Physical Review E 85~(2 Pt 1) (2012) 021906.

\bibitem{Zunino2017}
L.~Zunino, F.~Olivares, F.~Scholkmann, O.~A. Rosso, Permutation entropy based
  time series analysis: Equalities in the input signal can lead to false
  conclusions, Physics Letters A 381~(22) (2017) 1883--1892.

\bibitem{Yao2020APL}
W.~Yao, W.~Yao, D.~Yao, D.~Guo, J.~Wang, Shannon entropy and quantitative time
  irreversibility for different and even contradictory aspects of complex
  systems, Applied Physics Letters 116~(1) (2020) 014101.

\bibitem{Yao2019Sym}
W.~Yao, W.~Yao, J.~Wang, J.~Dai, Quantifying time irreversibility using
  probabilistic differences between symmetric permutations, Physics Letters A
  383~(8) (2019) 738--743.

\bibitem{Yao2022PLA}
W.~Yao, W.~Yao, J.~Wang, Comparative analysis of the original and amplitude
  permutations, Physics Letters A 430 (2022) 127977.

\bibitem{Zou2019}
Y.~Zou, R.~V. Donner, N.~Marwan, J.~F. Donges, J.~Kurths, Complex network
  approaches to nonlinear time series analysis, Physics Reports 787 (2019)
  1--97.

\bibitem{Pessa2020}
A.~A. Pessa, H.~V. Ribeiro, Mapping images into ordinal networks, Physical
  Review E 102~(5) (2020) 052312.

\bibitem{Kelly1979}
F.~P. Kelly, Reversibility and stochastic networks, Cambridge University Press,
  1979.

\bibitem{Diks1995}
C.~Diks, J.~C. Van~Houwelingen, F.~Takens, J.~Degoede, Reversibility as a
  criterion for discriminating time series, Physics Letters A 201~(2–3)
  (1995) 221--228.

\bibitem{Ramsey1995}
J.~B. Ramsey, P.~Rothman, Time irreversibility and business cycle asymmetry,
  Journal of Money Credit \& Banking 28~(1) (1995) 1--21.

\bibitem{Evans1994}
D.~J. Evans, D.~J. Searles, Equilibrium microstates which generate second law
  violating steady states, Physical Review E 50~(2) (1994) 1645--1648.

\bibitem{Evans2002}
D.~J. Evans, D.~J. Searles, The fluctuation theorem, Advances in Physics 51~(7)
  (2002) 1529--1585.

\bibitem{Searles2004}
D.~J. Searles, D.~J. Evans, Fluctuations relations for nonequilibrium systems,
  Australian Journal of Chemistry 57 (2004) 1119--1123.

\bibitem{Sevick2008}
E.~M. Sevick, R.~Prabhakar, S.~R. Williams, D.~J. Searles, Fluctuation
  theorems, Annual Review of Physical Chemistry 59~(1) (2008) 603--633.

\bibitem{Evans2016}
D.~J. Evans, D.~J. Searles, S.~R. Williams, Fundamentals of classical
  statistical thermodynamics: dissipation, relaxation, and fluctuation
  theorems, John Wiley \& Sons, 2016.

\bibitem{Seife2012}
U.~Seifert, Stochastic thermodynamics, fluctuation theorems and molecular
  machines, Reports on Progress in Physics 75~(12) (2012) 126001.

\bibitem{Yao2021DES}
W.~Yao, W.~Yao, Y.~Ju, Y.~Xia, D.~Guo, D.~Yao, Distribution of equal states for
  amplitude fluctuations in epileptic eeg, Biomedical Signal Processing and
  Control 69 (2021) 102738.

\bibitem{Schrei1996}
T.~Schreiber, A.~Schmitz, Improved surrogate data for nonlinearity tests,
  Physical Review Letters 77~(4) (1996) 635--638.

\bibitem{Schreiber2000}
T.~Schreiber, A.~Schmitz, Surrogate time series, Physica D: Nonlinear Phenomena
  142~(3–4) (2000) 346--382.

\bibitem{Stefanov2020}
A.~Stefanovska, P.~V. McClintock, Physics of Biological Oscillators: New
  Insights Into Non-Equilibrium and Non-Autonomous Systems, Springer Nature,
  2020.

\bibitem{epilepsy}
Epilepsy, \url{https://www.who.int/news-room/fact-sheets/detail/epilepsy/}.

\bibitem{Moshe2015}
S.~L. Moshé, E.~Perucca, P.~Ryvlin, T.~Tomson, Epilepsy: new advances, The
  Lancet 385~(9971) (2015) 884--898.

\bibitem{Lehner2014}
K.~Lehnertz, G.~Ansmann, S.~Bialonski, H.~Dickten, C.~Geier, S.~Porz, Evolving
  networks in the human epileptic brain, Physica D: Nonlinear Phenomena 267~(1)
  (2014) 7--15.

\bibitem{Schindle2016}
K.~Schindler, C.~Rummel, R.~G. Andrzejak, M.~Goodfellow, F.~Zubler, E.~Abela,
  R.~Wiest, C.~Pollo, A.~Steimer, H.~Gast, Ictal time-irreversible intracranial
  eeg signals as markers of the epileptogenic zone, Clinical neurophysiology
  127~(9) (2016) 3051--3058.

\bibitem{Andrze2001}
R.~G. Andrzejak, K.~Lehnertz, F.~Mormann, C.~Rieke, P.~David, C.~E. Elger,
  Indications of nonlinear deterministic and finite-dimensional structures in
  time series of brain electrical activity: dependence on recording region and
  brain state, Physical Review E 64~(6 Pt 1) (2001) 061907.

\bibitem{Iasemi2003}
L.~D. Iasemidis, Epileptic seizure prediction and control, IEEE Transactions on
  Biomedical Engineering 50~(5) (2003) 549--558.

\bibitem{Cui2019}
Y.~Cui, J.~Liu, Y.~Luo, S.~He, Y.~Xia, Y.~Zhang, D.~Yao, D.~Guo, Aberrant
  connectivity during pilocarpine-induced status epilepticus, International
  Journal of Neural Systems 30~(5) (2019) 1950029.

\bibitem{Kulp2017}
C.~W. Kulp, L.~Zunino, T.~Osborne, B.~Zawadzki, Using missing ordinal patterns
  to detect nonlinearity in time series data, Physical Review E 96~(2) (2017)
  022218.

\bibitem{Yao2021CSF}
W.~Yao, W.~Yao, J.~Wang, A novel parameter for nonequilibrium analysis in
  reconstructed state spaces, Chaos, Solitons \& Fractals 153 (2021) 111568.

\bibitem{David2020}
D.~Cuesta~Frau, Using the information provided by forbidden ordinal patterns in
  permutation entropy to reinforce time series discrimination capabilities,
  Entropy 22~(5) (2020) 494.

\bibitem{Amigo2008}
J.~M. Amigo, S.~Zambrano, M.~A. Sanjuán, Combinatorial detection of
  determinism in noisy time series, Europhysics Letters 83~(6) (2008) 60005.

\bibitem{Crooks1999}
G.~E. Crooks, Entropy production fluctuation theorem and the nonequilibrium
  work relation for free energy differences, Physical Review E 60~(3) (1999)
  2721.

\bibitem{Kim1999}
H.~S. Kim, R.~Eykholt, J.~D. Salas, Nonlinear dynamics, delay times, and
  embedding windows, Physica D: Nonlinear Phenomena 127 (1999) 48--60.

\bibitem{Kugium1996}
D.~Kugiumtzis, State space reconstruction parameters in the analysis of chaotic
  time series — the role of the time window length, Physica D: Nonlinear
  Phenomena 95~(1) (1996) 13--28.

\bibitem{Costa2002}
M.~D. Costa, A.~L. Goldberger, C.~K. Peng, Multiscale entropy analysis of
  complex physiologic time series, Physical Review Letters 89~(6) (2002)
  068102.

\end{thebibliography}

\end{document}